\documentstyle[preprint,aps]{revtex}
\begin{document}
\draft
\title
{Solitonic excitations in the Haldane phase of a S=1 chain}
\author{G.\ F\'ath and J.\ S\'olyom}
\address{Research Institute for Solid State Physics \\
               H-1525 Budapest, P.\ O.\ Box 49, Hungary}
\date{\today}
\maketitle
\begin{abstract}
We study low-lying excitations in the 1D $S=1$ antiferromagnetic
valence-bond-solid (VBS) model. In a numerical calculation on
finite systems the lowest excitations are found to form a
discrete triplet branch, separated from the higher-lying
continuum.  The dispersion of these triplet excitations can be
satisfactorily reproduced by assuming approximate wave functions.
These wave functions are shown to correspond to moving hidden
domain walls, i.e. to one-soliton excitations.
 \end{abstract}

\narrowtext
\section{Introduction}

 It was almost a decade ago that Haldane \cite{Ha} conjectured the
existence of a new type of ground state for isotropic Heisenberg
antiferromagnets (HAF) of integer spin $S$. The Haldane phase was
proposed to be characterized by a unique, disordered ground state
with exponential decay of the correlation functions and a finite
energy gap to the excited states. Since then
one-dimensional quantum spin chains with $S=1$ have been studied
intensively and it is claimed that the gapful behavior is a generic
feature of integer-spin models \cite{Af}.

The first rigorous example of a $S=1$ antiferromagnetic model with
Haldane phase was given by Affleck {\it et al.} \cite{Af-etal}.
They showed that the $S=1$ isotropic bilinear-biquadratic model
defined by the Hamiltonian
\begin{eqnarray}
H &=& \sum_{j=1}^N h_j(\beta)
  = \sum_{j=1}^N \left[ {1\over 2}{\bf S}_{j}\cdot {\bf S}_{j+1}
  +{\beta\over 2}\; ({\bf S}_{j}\cdot {\bf S}_{j+1})^2
  +{1\over 3} \right]\;,                              \label{eq:Ham}
\end{eqnarray}
has a short range valence-bond-solid ground state
for $\beta=1/3$ (VBS or AKLT model). At that
point $h_j(1/3)$ is a special projection operator, that projects out
the quintuplet state of the two neighboring spins, and it is positive
semi-definite. Therefore, a state $\Omega$ for which $h_j\Omega =0$
for any $j$ is necessarily a ground state with ground-state energy
$E_{\rm GS}=0$.  Such an $\Omega$ state could be constructed using
nearest-neighbor valence bonds. They were also able to prove rigorously
\cite{Af-etal} that in the infinite chain limit this state is the
only ground state, it is separated by a finite gap from the excited
states and the two-point correlation functions decay
exponentially.

According to Haldane's conjecture such a phase should not appear for
half-integer values of the spin. This was proven rigorously by
Affleck and Lieb \cite{Af-Li} and independently by Kolb \cite{Ko}.
It was shown for a wide class of models that
in the case when the ground state is a spin
singlet, the energy spectrum as a function of momentum $k$ is
symmetric under reflections with respect to $k=m\pi/2$ ($m$ integer),
and therefore the ground state should be at least doubly degenerate.

A similar proof fails in the integer $S$ case, allowing for the
existence of a unique singlet ground state \cite{Af-Li,Ko}. The
excitation spectrum is in general symmetric with respect to
$k=m\pi$ only. Of course, higher symmetry can also appear in
integer $S$ chains, as e.g.\ in the spontaneously dimerized
phase of the general bilinear-biquadratic $S=1$ model \cite{Fa-So}.

The $S=1$ bilinear-biquadratic model is integrable \cite{Fa-Ta} at the
critical point $\beta =1$,  that separates the dimerized phase
from the Haldane phase. Although the spin is integer, at this point the
excitations can be described in exactly the same
way as for the spin-1/2 HAF. More generally it is known since the
work by Faddeev and Takhtajan \cite{Fa-Ta}, that there are
integrable spin models for arbitrary $S$ in which the elementary
excitations are in fact spin-1/2 solitons with a dispersion independent
of the spin length $S$. The observable excitations are composite
particles, since due to topological reasons the solitons can appear
in singlet or triplet pairs only. As the energy of such a soliton
pair can be described by two
parameters, the excitations form a continuum in momentum space.

Away from the integrable point, where the symmetry properties of
the excitation spectra are different for integer and half-integer $S$,
the above mentioned picture of composite excitations may not hold.
In this paper we will study this problem.

We will restrict ourselves to the $S=1$ case only, where the
non-integrability appears in the most dramatic way in the Haldane phase.
We will show that the lowest excitations are real spin-1, one-particle
excitations, they cannot be decomposed into pairs of spin-1/2 solitons.
These triplet excitations are, however, not usual
antiferromagnons but rather some sort of
{\em hidden spin-1 solitons\/}.

The solitonic nature of the excitations of the integer spin models was
predicted by Haldane already. This assumption became less astonishing
after the discovery of the hidden
order in the Haldane phase. Recently in a very inspiring work, den Nijs
and Rommelse \cite{dN-Ro} have introduced a nonlocal
string operator $\sigma_{ij}^{\alpha}$ defined by
\begin{equation}
\sigma_{ij}^{\alpha}:=-S_i^{\alpha} \exp\left[
  i\pi\sum_{l=i+1}^{j-1}S_l^{\alpha}\right]S_j^{\alpha} \;.
\end{equation}

They argued that although in the Haldane phase the ground state
is disordered in the conventional sense, it has a hidden long-range
order that could be characterized by the string order parameter
\begin{equation}
O^{\alpha}_{\rm string}(H)=
\lim_{|i-j|\to\infty} {}_H\langle \sigma_{ij}^{\alpha} \rangle _H\;,
                  \quad \alpha=x,y,z,
      \label{sigma}
\end{equation}
where ${}_H\langle\cdot\rangle _H$ denotes the expectation value in the
ground state of the Hamiltonian $H$.

This prediction was later verified numerically by several
authors \cite{hidden}. The appearance of the hidden long-range order was
further discussed by Kennedy and Tasaki \cite{Ke-Ta}. They showed,
using a nonlocal unitary transformation, that
$O^{\alpha}_{\rm string}>0$ corresponds to the spontaneous breaking of a
hidden $Z_2\times Z_2$ symmetry of the model.
Similarly, the fact that the four lowest states of an open chain are
exponentially close to each other is also a consequence of this
broken symmetry.
It is generally expected, that the breaking of a discrete symmetry
in the ground state leads to an excitation gap, since Goldstone
bosons do not appear. Excitations of the model can then be thought of as
some sort of (hidden) domain walls, separating regions with
different ground states. This picture was made more explicit by
Elstner and Mikeska \cite{El-Mi}, who used spin-zero defects \cite{Go}
to disorder the antiferromagnetic state. The spin-zero defects are in
fact solitons. One of the main goals of this paper is to further
examine this problem.

We will use numerical and analytical methods to study the low-lying
excitations in the $\beta=1/3$ case. Beside the fact that the ground state
of the VBS model can be constructed analytically, there is another
good reason to focus on this model.
In a recent study of the general bilinear-biquadratic model of Eq.\
(\ref{eq:Ham}), we observed \cite{Fa-So} that the convergence of
various finite-size estimates to their thermodynamic limit is
extremely fast in the close vicinity of $\beta=1/3$.
This is certainly not true for general $\beta$. Moving away from
$\beta=1/3$ finite-size corrections become stronger, and one must
consider longer and longer chains in order to see the real asymptotic
behaviour. The rapid convergence at the VBS point may not
be very surprising, if we remember, that at this point in the ground state
first-neighbor valence bonds are only present and the
ground-state energy density becomes independent of the chain length.
Although the excited states do show some dependence on $N$, this is
found to be exponentially small for the most relevant levels. Therefore
extrapolation from finite-size calculations allows to draw quite reliable
conclusions on the spectrum and it can be compared directly to the
analytical (variational) results.

After a detailed numerical analysis of the spectrum, we will study
trial wave functions for the elementary excitations and
illustrate their solitonic nature. Since the Haldane phase at $\beta =0$ is
believed to be in the same universality class as the VBS model
at $\beta =1/3$,
our finding should be qualitatively correct for the usual HAF.

The layout of the paper is as follows: In Sec.\ II we present our
numerical results. Section III contains the analysis of two seemingly
different trial wave functions and their equivalence is shown. The
elementary excitations are argued to be triplet bonds in the VBS
structure. In Sec.\ IV, we recall the nonlocal unitary transformation
of Kennedy and Tasaki. The trial wave functions for the excitations
are studied further, using this transformation, in Sec.\ V.
Thus their domain wall nature becomes explicit. Finally, in Sec.\ VI we
summarize our results.

\section{Numerical Results}

Our first aim is to study the excitation spectrum of the
VBS model numerically, using a periodic boundary condition. The
symmetry properties of the Hamiltonian in Eq.\ (\ref{eq:Ham}) allow
us to classify the eigenstates according to their total spin
$S_T$, its component along the $z$-axis $S_T^z$, and the momentum
$k=2\pi l/N$ ($l$ integer) of the states, where N is the length of
the chain. We computed several low-lying eigenvalues of the
Hamiltonian for each possible value of $k$, using a L\'anczos
algorithm, and also determined the total spin of the states.  Chains with
an even number of sites up to $N=16$ were considered.

Figure \ref{spec14} shows the six lowest eigenvalues for all $k$ in
our longest chain with $N=16$. For some of the
energies the total spin quantum number $S_T$ is also given. It is seen
that in full agreement with all previous results \cite{Af}, the
lowest excited state (denoted by $A$) is an $S_T=1$ state with
momentum $k=\pi$. Moreover, also in the whole range $|k|\agt\pi/2$
the lowest-energy excited states have the same total spin
$S_T=1$. In the thermodynamic limit, these states, as a function of $k$,
seem to form a continuous
branch of excitations. In fact, according to the general theorem we
proved in Ref.\ \cite{Fa-So},  an excitation with
$S_T\ge 1$ cannot remain isolated in the $N\to\infty$ limit.

On the other hand, near $k=\pi$, the energies of the next higher
lying excitations are situated at a distance from the triplet
branch, that is much greater than their average energy difference
from each other. This behaviour indicates that in the infinite
chain limit, at least near $k=\pi$, the lowest triplet excitations
do not belong to a continuum. The existence of such a discrete
branch below the higher lying continuum seems to be another
characteristic feature of the non-integrable integer-spin models.
Unlike the integrable spin models and  general spin models of
half-integer $S$, the lowest-lying excitations in the Haldane phase
are real one-particle spin-1 excitations. They cannot be decomposed
into $S=1/2$ solitons.

Above the triplet branch, the higher-lying excitations of the
VBS model seem to be ``dense'' for all $k$, supposedly forming a
continuum. They probably cannot be described by a single parameter.
For $N=16$, the triplet branch merges into this continuum
somewhere below $k\sim\pi/2$. Near $k=0$ there is no sign of
a discrete branch, here the lowest-lying excitations are thought
to belong to the continuum.

There is another remarkable feature that can be observed in our
finite-chain calculation. The gap $\Delta_B$ to the lowest
excited state of the $k=0$ subspace is approximately twice the
singlet-triplet gap $\Delta_A$ at $k=\pi$. The same property was
observed by Takahashi \cite{Ta} for the pure Heisenberg chain
$\beta=0$.  Similarly, the gap $\Delta_C$ between the ground state
and the second excited state in the $k=\pi$ sector seems to be three
times as large as the singlet-triplet gap. It is also noteworthy that
state $B$ (the lowest $k=0$ excited state) is a quintuplet ($S_T=2$)
state and $C$ (the second lowest $k=\pi$ excited state) is a state
with $S_T=3$.  The
physical picture behind such a behavior is simple. The excitations
near $k=0$ can be composed of two low-lying excitations near $k=\pi$
and similarly, three excitations near $k=\pi$ can be combined to give
another excitation near $k=\pi$.

Whether the spectrum has this property in the $N\to\infty$ limit
was tested by extrapolating the finite-size calculations to
infinitely long chains. In Fig.\ \ref{fss} the
gaps $\Delta_A$, $\Delta_B$, and $\Delta_C$ are plotted as a
function of $1/N$. The convergence to the thermodynamic limit is
very fast, especially for $\Delta_A$. Using standard extrapolation
methods, the limiting values of the three quantities are:
$\Delta_A=0.350\,124\pm 10^{-6}$, $\Delta_B=0.71\pm 0.01$,
and $\Delta_C=1.09\pm 0.03$,
respectively. As is seen, the ratios give the anticipated values
$1:2:3$ within 4\% of error.

This numerical calculation supports rather convincingly the idea that
the elementary excitations of the model form a discrete triplet
branch, which is separated from the multi-particle continuum
in a wide range around $k=\pi$. Analyzing the lower boundary of
this multi-particle continuum, it seems very likely that the
two-particle states
%%%(or much rather two-{\it soliton} states, as we will show soon)
near $k=0$ are essentially {\em scattering} states of two elementary
excitations. The energy and momentum of such a multi-particle state is
then simply the sum of the energies and momenta,
respectively, of the two particles. The situation is
similar for the three-particle states near $k=\pi$. In the numerical
calculation there does not seem to be any sign of bound states
below the scattering continuum.

\section{Trial wave functions for the elementary excitations}

In this Section we will study the elementary excitations of the VBS
model analytically. We will see that it is possible to reproduce
the dispersion relation of the discrete triplet branch quite precisely
by assuming simple trial wave functions. It will be argued that the
elementary excitations are (hidden) solitons that destroy the
hidden order of the ground state.

First we recall the form of the ground-state wave function.
As it was mentioned in the Introduction, the ground state of the
VBS model can be constructed analytically using nearest
neighbour valence-bonds. For this an $S=1$ operator will be
composed of two $S=1/2$ operators. Taking the tensor product of the
two spin-1/2 spaces, a new orthogonal basis at site $i$ \cite{Af-etal}
is constructed in the form
\begin{equation}
\psi_{\alpha \beta}^i=[\psi_{\alpha} \otimes \psi_{\beta} +
        \psi_{\beta} \otimes \psi_{\alpha} ]/\sqrt{2}\;,
\end{equation}
where $\psi_{\alpha}$ and $\psi_{\beta}$ represent the eigenstates of
the two spin-1/2 operators, and the Greek indices take the values
$\uparrow$ and $\downarrow$. There are three independent symmetric
combinations corresponding to the three eigenstates of the spin-1
operator, $|+\rangle_i$, $|0\rangle_i$,
and $|-\rangle_i$ with $S^z_i=1, 0 $ and $-1$, respectively,
\begin{eqnarray}
|+\rangle_i&=&\psi_{\uparrow \uparrow}^i/ \sqrt{2}\;, ]\nonumber \\
|0\rangle_i &=&\psi_{\uparrow \downarrow}^i
        \equiv \psi_{\downarrow \uparrow}^i \label{psinorm}\;, \\
|-\rangle_i&=&\psi_{\downarrow \downarrow}^i / \sqrt{2}\;. \nonumber
\end{eqnarray}
The fourth orthogonal state that completes the basis is the
antisymmetric combination corresponding to an
$S=0$ state at site $i$. This configuration will be excluded.

The ground-state wave function of an open chain of length $N$ can
be written in terms of these states as \cite{Af-etal}
\begin{eqnarray}
\Omega(\alpha_1,\beta_N) &=&
\psi^1_{\alpha_1\beta_1}\varepsilon^{\beta_1\alpha_2}
     \psi^2_{\alpha_2 \beta_2}\varepsilon^{\beta_2\alpha_3}\dots
    \psi^i_{\alpha_i \beta_i}\varepsilon^{\beta_i \alpha_{i+1}}
    \dots \psi^N_{\alpha_N \beta_N}\;.
    \label{Omegaopen}
\end{eqnarray}
Here and in what follows summation is meant over repeated indices.
$\varepsilon^{\alpha \beta}$ is an antisymmetric tensor with
$\varepsilon^{\uparrow \downarrow}=-\varepsilon^{\downarrow \uparrow}
=1$. At both ends of the chain there is a loose spin-1/2 degree of freedom,
denoted by $\alpha_1$ and $\beta_N$. Since both can assume any of the
two eigenstates independently, the ground state is fourfold degenerate.
Three of these states constitute the three components of
a spin triplet, while the fourth state is a spin singlet.
It was shown, however, that these four ground states converge to the
same infinite volume limit as $N\to\infty$.

A unique ground state can be formed even for finite $N$ in the case of
periodic boundary condition by antisymmetrizing the two loose end spins.
The ground-state wave function can be written as
\begin{equation}
\Omega = \Omega(\alpha_1,\beta_N)\varepsilon^{\beta_N\alpha_{1}}\;.
    \label{Omega}
\end{equation}
Note that these states are not normalized, $||\Omega
(\alpha_1,\beta_N)||^2={1\over 2}3^N+{\cal O}(1)$ and $||\Omega
||^2=3^N+{\cal O}(1)$.

$\Omega$ has the interesting property that the configurations appearing
in $\Omega$ look like in the conventional $S^z$ representation as
\begin{equation}
 \dots 0+0\dots 0-0\dots 0+0\dots 0-0\dots\;, \label{gshidden}
\end{equation}
i.e., each $+$ is followed by a $-$ with an arbitrary number of 0
states in-between and vice versa. This is nothing but a dilute
spin-1/2 N\'eel antiferromagnet, where
the 0's represent a background and the $+$ and $-$ states denote
the two possible degrees of freedom of a spin-1/2 particle.
As is seen, the hidden N\'eel order is perfect for the VBS
ground state. For this case the string order parameter of den
Nijs and Rommelse is \cite{dN-Ro}
$O^{\alpha}_{\rm string}(H_{\rm VBS})={4\over 9}$, $\alpha=x,y,z$.
Moving away from the VBS point, quantum
fluctuations begin to destroy the above structure of $\Omega$.
However, the hidden long-range order, characterized by
$O^{\alpha}_{\rm string}>0$, $\alpha=x,y,z$, is expected to persist
in a wide region, in the whole Haldane phase
\cite{hidden}.

As for the excited states of the VBS model, our knowledge is
much less accurate, since the eigenfunctions cannot be
constructed in a similarly rigorous way. Recently, however, two
seemingly rather different trial wave functions were proposed to
describe elementary excitations in the model. Arovas, Auerbach and
Haldane \cite{ArAuHa} proposed the form
\begin{equation}
\vert k\rangle
        =N^{-1/2}\sum_{j=1}^N e^{i k j}S_j^{\mu} \vert\Omega\rangle\;,
        \quad \mu=z,+,-,                                \label{sma}
\end{equation}
and obtained
\begin{equation}
\epsilon(k)={\langle k|H_{\rm VBS}| k\rangle\over
\langle k|k \rangle}={25+15\cos(k)\over 27}\;,
\label{disp}
\end{equation}
for the dispersion relation of the excitations. As it is seen from
Fig.\ \ref{fss}, the variational ansatz yields an upper bound
$\Delta_A\le {10\over 27}=0.3704$, very close to the real excitation
gap obtained from the finite-size calculation at $k=\pi$.

The dispersion relation Eq.\ (\ref{disp}) is plotted in Fig.\ \ref{spec14}
with a dashed line. Comparison with the numerical results suggest that
the trial wave function $|k\rangle$ gives a reasonable
descriptions of the elementary excitations not only at $k=\pi$,
but in a large region of the Brillouin zone, in the range $|k|\agt \pi/2$.
Below that, in the region near $k=0$, the two-particle scattering continuum
dominates the spectrum.

Looking at the form of the trial wave function in Eq.\
(\ref{sma}), one is tempted to interpret the elementary
excitations as {\em magnons}. Note, however, that although these
"magnons" would not destroy a conventional long range order, they
do destroy the hidden order. This is directly seen if we use
e.g.\ the $\mu=+$ component in Eq.\ (\ref{sma}). In $S^+_j\Omega$
only those configurations appear that look like
\begin{equation}
\dots 0-0\dots 0+0 \dots 0+0\dots 0-0\dots 0+0\dots\;, \label{soli}
\end{equation}
i.e., the N'el order of the nonzero components is broken at one point.
In the dilute antiferromagnetic picture this is nothing else but a
usual antiferromagnetic soliton embedded in the
background of $0$'s. Such a soliton, unlike a magnon, destroys the long
range order, in this case the {\it hidden} antiferromagnetic order
corresponding to $O^z_{\rm string}$.

The solitonic nature of $S^z_j\Omega$ is much less
obvious at first sight, since in this case the hidden N\'eel order
of Eq.\ (\ref{gshidden}) seems to remain intact. However, as it will be
illustrated in Sec.\ V, now the hidden order in the transverse
directions (i.e.\ $O^x_{\rm string}$ and $O^y_{\rm string}$) will
be destroyed.

In an alternative approach Knabe \cite{Kna} proposed the following
construction for the elementary excitations. Let us retain
the valence-bond structure of $\Omega$ for every bond except
between sites $j$ and $j+1$, where the two spin-1/2 degrees of
freedom $\beta_j$ and $\alpha_{j+1}$ are symmetrized to form
a triplet bond, instead of the antisymmetrized singlet bond.

To make this construction more explicit, divide
the sites of the chain into two sets: ${\cal L}=\{1,2,\dots,j\}$
and ${\cal R}=\{j+1,j+2,\dots,N\}$. We define first
the states $\Phi_j^{\beta_j,\alpha_{j+1}}(\alpha_1,\beta_N)$,
with $\beta_j,\alpha_{j+1}=\uparrow,\downarrow$ and $1\le j<N$,
as explicit tensor products of two arbitrary ground states, one
on $\cal L$ and the other on $\cal R$, respectively, as
\begin{equation}
\Phi_j^{\beta_j,\alpha_{j+1}}(\alpha_1,\beta_N)=
\Omega_{\cal L}(\alpha_1,\beta_j)\otimes\Omega_{\cal R}
          (\alpha_{j+1},\beta_N)\;. \label{phifolle}
\end{equation}
Obviously the ground state of the full chain can be obtained by
antisymmetrizing with respect to  $\alpha_j$ and $\beta_{j+1}$, i.e., by
connecting the $\cal L$ and $\cal R$ sides with a singlet
valence bond,
\begin{equation}
\Omega(\alpha_1,\beta_N)=
                        \Phi_j^{\uparrow\downarrow}(\alpha_1,\beta_N)
                       -\Phi_j^{\downarrow\uparrow}(\alpha_1,\beta_N)\;.
\end{equation}
On the other hand, {\em symmetrization} with respect to $\alpha_j$ and
$\beta_{j+1}$ defines three new states,
\begin{eqnarray}
\Phi_j^+(\alpha_1,\beta_N)&=&\Phi_j^{\uparrow\uparrow}(\alpha_1,\beta_N)\;,
                                                \label{pbond}\\
\Phi_j^0\,(\alpha_1,\beta_N)&=&{1\over \sqrt{2}}\left[
             \Phi_j^{\uparrow\downarrow}(\alpha_1,\beta_N)
             +\Phi_j^{\downarrow\uparrow}(\alpha_1,\beta_N)\right]\;,\\
\Phi_j^-(\alpha_1,\beta_N)&=&
              \Phi_j^{\downarrow\downarrow}(\alpha_1,\beta_N)\;.
                                                \label{nbond}
\end{eqnarray}
In these states the singlet valence bond between sites $j$ and $j+1$ is
substituted by a triplet bond. This adds an extra spin-1 degree of
freedom to the two free spin-1/2 variables in the ground states of the
{\em open} chain. The factor ${1/ \sqrt{2}}$ is
introduced in $\Phi_j^0$ to ensure that the states have
the same norm, $||\Phi_j^a(\alpha_1,\beta_N)||^2={1\over 4}3^N+{\cal
O}(1)$, $a=+,0,-$, as $N\to\infty$.

In the case of periodic boundary condition, the two loose spin-1/2
degrees of freedom at the chain ends should again be contracted
with an $\varepsilon$ tensor. This defines the functions
\begin{eqnarray}
\Phi_j^{a}=\Phi_j^{a}(\alpha_1,\beta_N)\varepsilon^{\beta_N,\alpha_1}\;,
                        \quad a=+,0,-,
\end{eqnarray}
for $1\le j\le N-1$. The state $\Phi_N^a$, in which the spin-1 bond
connects the last and first sites, $N$ and $1$, is defined analogously.
Unlike the case of open boundary condition, now
the total spin of the three states $\Phi_j^a$, $a=+,0,-$,
is necessarily $S_T=1$ with $S_T^z=+1, 0, -1,$ respectively,
since all the other antisymmetrized bonds have zero spin.

These states have again some soliton-like nature, just
as in the other approach above. $\Phi_j^+$, e.g.,
can be explicitly written as
\begin{eqnarray}
 \Phi_j^+ &=& \dots \varepsilon^{\beta_{j-2} \alpha_{j-1}}
        \psi^{j-1}_{\alpha_{j-1} \beta_{j-1}}
        \varepsilon^{\beta_{j-1} \alpha_{j}}
        \psi^{j}_{\alpha_j \uparrow}
        \psi^{j+1}_{\uparrow \beta_{j+1}}
        \varepsilon^{\beta_{j+1} \alpha_{j+2}}
        \psi^{j+2}_{\alpha_{j+2} \beta_{j+2}}
        \varepsilon^{\beta_{j+2} \alpha_{j+3}} \dots \;.
        \label{eq:Psi}
\end{eqnarray}
In each nonzero configuration the subscripts $\beta_i$ and
$\alpha_{i+1}$ are antiparallel except for $i=j$, for which
$\beta_j=\alpha_{j+1} =\uparrow$. Transforming this into the standard
$S_z$ representation, the nearest nonzero spin states on the left
and right side of the triplet bond are necessarily $+$ states. Otherwise
the N'el order of the $+$ and $-$ states is complete on both sides.

As a variational ansatz, Knabe \cite{Kna} analyzed a general linear
combination of the $\Phi_j^a$ states, $\sum_{j=1}^N c_j
\Phi_j^a$, and found that the energy is minimized if $c_j=(-1)^j$.
For the primary gap of the
model he obtained an upper bound in the form $\Delta_A\le {5\over
14}\approx 0.3571$. Correcting a small obvious mistake in the numerics of
his paper, the correct upper bound is $10/27\approx 0.3704$, exactly
as in Ref.\ \cite{ArAuHa}.

It is quite straightforward to generalize this calculation to arbitrary
momentum $k$. Looking for translationally invariant trial wave
functions, we define
\begin{equation}
  \Phi^a (k)=\sum_{j=1}^N e^{ikj} \Phi^a_j\;, \quad a=+,0,-.\label{transl}
\end{equation}
Knabe's wave function corresponds to $k=\pi$. With our definition of the
(unnormalized) trial wave functions $\Phi_j^a$, $a=+,0,-$, Knabe's
results can be reexpressed in the following form
\begin{equation}
\langle\Phi^a_j|\Phi^a_{j'}\rangle
  ={1\over 2} 3^N (-{1\over 3})^{|j-j'|}+{\cal O}(1)\;,
        \label{norm}
\end{equation}
and
\begin{equation}
  \langle\Phi^a_j|H_{\rm VBS}| \Phi^a_{j\prime}\rangle
  =\delta_{jj\prime} {10\over 27} 3^N+{\cal O}(1)\;.
                                    \label{PhiHPhi}
\end{equation}
Using these results, it is now straightforward to calculate the
normalization of the translationally invariant states
$\Phi^a(k)$ defined by Eq.\ (\ref{transl}), and
the expectation value of the energy in these states.
In the thermodynamic limit we get
\begin{eqnarray}
\langle\Phi^a(k)|\Phi^a(k)\rangle &=&
\sum_{j,j'=0}^N e^{ik(j'-j)} \langle\Phi^a_j|\Phi^a_{j'}\rangle
                                                      \nonumber\\
&=&N {1\over 2}3^N \sum_{r=0}^N e^{ikr}(-{1\over 3})^{r}
 = {2\over 5+3 \cos(k)} N 3^N,
\end{eqnarray}
and
\begin{eqnarray}
\langle\Phi^a(k)|H_{\rm VBS}| \Phi^a(k)\rangle &=&
\sum_{j,j'=0}^N e^{ik(j'-j)} \langle\Phi^a_j|H_{\rm VBS}|
\Phi^a_{j'}\rangle = {10\over 27} N 3^N.
\end{eqnarray}
Whence the dispersion of the excitations is again
\begin{equation}
    \epsilon(k)=
 {\langle\Phi^a(k)|H_{\rm VBS}| \Phi^a(k)\rangle\over
 \langle\Phi^a(k)|\Phi^a(k)\rangle}={25+15\cos(k)\over 27}\;.
\end{equation}

This dispersion is exactly the same as that obtained by Arovas {\em et al.}
in Eq.\ (\ref{sma}). Despite the different forms of the wave functions,
the identical result for the dispersion relation indicates a deep
connection between the two approximations. In fact, it is not too
difficult to show that $S_j^{\mu} \vert\Omega\rangle$ can be expressed
in a simple form with our $\Phi_j^{a}$ configurations:
\begin{eqnarray}
S_j^z \vert\Omega\rangle &=&{1\over \sqrt{2}}(\Phi_j^0-\Phi_{j-1}^0)\;,
                        \label{equiv1}\\
S_j^{\pm} \vert\Omega\rangle &=&\mp (\Phi_j^{\pm}-\Phi_{j-1}^{\pm})\;,
                \label{equiv2}
\end{eqnarray}
therefore $\vert k\rangle$ differs from $\Phi^{\mu}(k)$ in
a constant factor only, which is cancelled when
the expectation value of the energy is taken.

One may ask the question, which construction of the above two
should now be considered as the elementary excitation.
Eqs.\ (\ref{equiv1}) and (\ref{equiv2})
show that $S_j^{\mu} \vert\Omega\rangle$ is a simple linear
combination of the $\Phi_j^a$ configurations. The revers, however,
is not true. $\Phi_j^a$ cannot be expressed with  $S_j^{\mu} \vert
\Omega\rangle$ in a similarly simple way. Therefore, one
should conclude that the elementary excitations are in fact the moving
triplet bonds.

\section{The Kennedy-Tasaki transformation}

A better picture of the above described elementary  excitations of
the VBS model can be obtained by using the nonlocal unitary transformation
$U$ of Kennedy and Tasaki \cite{Ke-Ta}. This transforms the
antiferromagnetic Hamiltonian into a ferromagnetic-like model and
makes the $Z_2\times Z_2$ symmetry breaking in the ground state
explicit. First we recall some features of this transformation, then
show that the above trial
wave functions transform under $U$ into simple, explicit domain
walls. A small inconvenience, arising from the nonlocal
character of the transformation, is that it can only be used conveniently
on chains with {\em open} boundary conditions. Note, however, that
we do not seek for exact solutions but for variational results only, so
the boundary condition will have no relevance for long enough
chains.

We define the unitary $U$ in the usual way \cite{Ke-Ta}:
Let $|s\rangle=|s_1,s_2,\dots,s_N\rangle$ denote a basis state in the
$S^z$ representation, where $s_i=+,0,-$ stands for the eigenvalues
$+1, 0, -1$, respectively of $S^z_i$. Introducing new variables by
\begin{equation}
\overline{s}_i=\exp \left[ i\pi\sum_{l=1}^{i-1} s_l\right] s_i\;,
      \label{sbar}
\end{equation}
the spin configurations can be given as
$|\overline{s}\rangle=|\overline{s}_1,\overline{s}_2,\dots,
\overline{s}_N\rangle$. Note that in $|\overline{s}\rangle$ all the 0's of
$|s\rangle$ remain unchanged, while a $+$ or $-$ at site $i$ is flipped
or remains unchanged, depending on whether the number of $+$'s and $-$'s on
sites $1\le l<i$ is odd or even. The unitary $U$ is then defined by
\begin{equation}
U|s\rangle=(-1)^{M(s)}|\overline{s}\rangle\;,
\end{equation}
where $M(s)$ denotes the number of odd sites $i$ on which $s_i=0$.

Let us consider now the VBS Hamiltonian. It can be shown \cite{Ke-Ta}
that $h_j$, $j=1,\dots,N-1$, transforms under $U$ in a relatively
simple (local) way into
\begin{equation}
\tilde{h}_j=Uh_jU^{-1}=
               \matrix{\vert &\!\!\!\!\!\!++\/ \rangle \cr
                       \vert &\!\!\!\!\!\!+0\/ \rangle \cr
                       \vert &\!\!\!\!\!\!0+\/ \rangle \cr
                       \vert &\!\!\!\!\!\!+-\/ \rangle \cr
                       \vert &\!\!\!\!\! 0\,0\;\/ \rangle\; \cr
                       \vert &\!\!\!\!\!\!-+\/ \rangle \cr
                       \vert &\!\!\!\!\!\!0-\/ \rangle \cr
                       \vert &\!\!\!\!\!\!-0\/ \rangle \cr
                       \vert &\!\!\!\!\!\!--\/ \rangle }
{1\over 6} \pmatrix{ 1  &    &    &    &-2  &    &    &    & 1  \cr
                        & 3  &-3  &    &    &    &    &    &    \cr
                        &-3  & 3  &    &    &    &    &    &    \cr
                        &    &    &  6 &    &    &    &    &    \cr
                    -2  &    &    &    & 4  &    &    &    &-2  \cr
                        &    &    &    &    &  6 &    &    &    \cr
                        &    &    &    &    &    & 3  &-3  &    \cr
                        &    &    &    &    &    &-3  & 3  &    \cr
                     1  &    &    &    &-2  &    &    &    & 1  }\;.
\end{equation}
This, however, does not hold for $h_N$, i.e.\ for the term that
couples the last and the first spins of the chain.
$\tilde{h}_N=U h_N U^{-1}$ cannot be written in a similar
form, moreover it does not remain local either. This problem can,
however, be avoided if we switch to open boundary conditions. In this case
$\tilde{H}_{\rm VBS}$ simply reads as
\begin{equation}
\tilde{H}_{\rm VBS}=U H_{\rm VBS}U^{-1}=\sum_{j=1}^{N-1}
                                  \tilde{h}_j\;.
\end{equation}

The diagonalization of the above two-site Hamiltonian
$\tilde{h}_j$ shows that its ground-state
sector with zero energy is four dimensional, and is spanned by the states
$\phi_{\nu}\otimes \phi_{\nu}$, $\nu=1,2,3,4$ (here no summation is meant
over $\nu$), where the single-site states are
\begin{eqnarray}
\phi_1=(|0\rangle +\sqrt{2}|+\rangle )/\sqrt{3}\;,\\
\phi_2=(|0\rangle -\sqrt{2}|+\rangle )/\sqrt{3}\;,\\
\phi_3=(|0\rangle +\sqrt{2}|-\rangle )/\sqrt{3}\;,\\
\phi_4=(|0\rangle -\sqrt{2}|-\rangle )/\sqrt{3}\;.
\end{eqnarray}
Note, that this basis is not orthogonal, since $|\langle\phi_{\nu '}|
\phi_{\nu}\rangle|=1/3$, $\nu'\ne\nu$.

The ground states of an open chain with $N$ sites can simply be written
as the tensor product of the above introduced single-site states,
\begin{equation}
\Psi_{\nu}= \phi^1_{\nu}\otimes\phi^2_{\nu}\otimes\cdots\phi^{N-1}_{\nu}
       \otimes\phi^N_{\nu}\;, \quad {\nu}=1,2,3,4. \label{fourGS}
\end{equation}
These wave functions and the ground-state wave functions described
in Eq.\ (\ref{Omegaopen}) can be easily related.
On an $L$-site lattice any linear combinations of the
$\Psi_{\nu}$'s are ground states. Such a linear combination is
obtained if $U$ acts directly on the ground state
$\Omega(\alpha_1,\beta_L)$ of $H_{\rm VBS}$. For all the
configurations in $\Omega(\alpha_1,\beta_L)$, the value of
$\alpha_1$ decides unequivocally the sign of the first nonzero spin
(if $\alpha_1=\uparrow$ then it should be a $+$), and then for a given
value of $L$, $\beta_L$ fixes the parity of the total number of
nonzero spins. Using these two observations, the following relations
can easily be worked out
\begin{eqnarray}
U\Omega(\uparrow,\uparrow)&=&
  (-1)^{1+L+[L/2]}(3)^{L/2}{1\over 2}(\Psi_1-\Psi_2)\;,
                                                \label{Oleft}\\
U\Omega(\uparrow,\downarrow)&=&
  (-1)^{1+[L/2]}  (3)^{L/2}{1\over 2}(\Psi_1+\Psi_2)\;,\\
U\Omega(\downarrow,\uparrow)&=&
  (-1)^{L+[L/2]}  (3)^{L/2}{1\over 2}(\Psi_3+\Psi_4)\;,\\
U\Omega(\downarrow,\downarrow)&=&
  (-1)^{[L/2]}    (3)^{L/2}{1\over 2}(\Psi_3-\Psi_4)\;,
                                                \label{Oright}
\end{eqnarray}
where $[L/2]$ denotes the integer part of $L/2$.

It can be shown rigorously that the four states in Eq.\ (\ref{fourGS})
remain the only ground states as $N\to\infty$,
and they converge to four {\em different}
infinite volume ground states. Note that the ground-state
degeneracy of the original and the transformed Hamiltonians differ
in the infinite volume limit. This is a consequence of the
nonlocality of $U$.

The states in Eq.\ (\ref{fourGS}) have
long-range order reflecting the spontaneous breaking of a
$Z_2\times Z_2$ symmetry, the only explicit (local) symmetry of
$\tilde{H}_{\rm VBS}$. Introducing the ferromagnetic order
parameter of the transformed system
\begin{equation}
O^{\alpha}_{\rm ferro}(\tilde{H})=
\lim_{|i-j|\to\infty} {}_{\tilde{H}}\langle
             S_i^{\alpha}S_j^{\alpha} \rangle _{\tilde{H}}\;,
             \quad \alpha=x,y,z,              \label{op}
\end{equation}
one obtains that
\begin{equation}
O^{\alpha}_{\rm ferro}(\tilde{H})=
{}_{\tilde{H}}\langle S_i^{\alpha}\rangle _{\tilde{H}}^2 ={4\over 9}\;,
             \quad \alpha=x,z.
\end{equation}
 While ${}_{\tilde{H}}\langle S_i^z\rangle _{\tilde{H}}=+{2\over 3}$ for
$\Psi_1$ and $\Psi_2$, we find
${}_{\tilde{H}}\langle S_i^z\rangle _{\tilde{H}}=-{2\over 3}$
for $\Psi_3$ and $\Psi_4$. Similarly,
${}_{\tilde{H}}\langle S_i^x\rangle _{\tilde{H}}=+{2\over 3}$
for $\Psi_1$ and $\Psi_3$, and
${}_{\tilde{H}}\langle S_i^x\rangle _{\tilde{H}}=-{2\over 3}$
for $\Psi_2$ and $\Psi_4$.
The appearance of the long-range order in
the transformed Hamiltonian corresponds to a non-vanishing value of the
string order parameter in the original system, since by the equivalence
\begin{equation}
O^{\alpha}_{\rm string}(H)=
O^{\alpha}_{\rm ferro}(\tilde{H})\quad \mbox{for}\quad
\alpha=x,z,
\end{equation}
the string order transforms into a ferromagnetic order under
$U$ \cite{Ke-Ta}. This equivalence does not hold for the $y$
component.

\section{Solitons in the Kennedy-Tasaki transformation}

Our aim now is to show how our soliton configurations $\Phi^a_j$,
$a=+,0,-$ transform under the unitary transformation. It will be
found that in the transformed model they are explicit domain walls
separating regions with different ground states $\Psi_{\nu}$. Since
we work now with open boundary conditions, the two loose spin-1/2
variables at the left and right chain ends should be retained
explicitly.

For this purpose, we cut the chain into ${\cal L}$ and ${\cal R}$ parts
as in Section III, and define new unitary operators.
Since a configuration $|s\rangle$ can be written as an explicit tensor
product of the left and right states,
\begin{equation}
|s_1,s_2,\dots,s_N\rangle=
|s_1,s_2,\dots,s_j\rangle \otimes
                          |s_{j+1},s_{j+2},\dots,s_N\rangle\;.
\end{equation}
the action of $U$ on $|s\rangle$ can be given in the form
\begin{equation}
U|s_1,s_2,\dots,s_N\rangle=
U_j^{\cal L}|s_1,s_2,\dots,s_j\rangle \otimes
U_j^{\cal R}|s_{j+1},s_{j+2},\dots,s_N\rangle\;. \label{Uaction}
\end{equation}
Here
\begin{eqnarray}
U_j^{\cal L}|s_1,s_2,\dots,s_j\rangle &=&
(-1)^{M_{\cal L}(s)}|\overline{s}_1,\overline{s}_2,\dots,
        \overline{s}_j\rangle\;,\\
U_j^{\cal R}|s_{j+1},s_{j+2},\dots,s_N\rangle &=&
(-1)^{M_{\cal R}(s)}|\overline{s}_{j+1},\overline{s}_{j+2},
        \dots,\overline{s}_N\rangle\;,
\end{eqnarray}
with $|\overline{s}\rangle$ defined by Eq.\ (\ref{sbar}). $M_{\cal
L}(s)$ ($M_{\cal R}(s)$) is the number of odd sites for which $s_i=0$ and
$i\in\cal L$ ($i\in\cal R$). Obviously $M_{\cal
L}(s)+M_{\cal R}(s)=M(s)$.

It is seen that in general the effect
of $U_j^{\cal R}$ on $|s_{j+1},s_{j+2},\dots,s_N\rangle$
cannot be calculated without some explicit knowledge of the
configuration on $\cal L$, since e.g.\ the first nonzero spin of $\cal
R$ is flipped according to whether the parity of the number of nonzero
spins of $\cal L$ is even or odd. Therefore, we define another
unitary $V_j^{\cal R}$, that will be independent of $\cal L$, in
the following way:
\begin{equation}
V_j^{\cal R}|s_{j+1},s_{j+2},\dots,s_N\rangle =
(-1)^{K_{\cal R}(s)}
|\overline{\overline{s}}_{j+1},\overline{\overline{s}}_{j+2},\dots,
\overline{\overline{s}}_N\rangle\;,
\end{equation}
where
\begin{equation}
\overline{\overline{s}}_i=
      \exp \left[ i\pi\sum_{l=j+1}^{i-1} s_l\right] s_i\;,
      \quad i\ge j+1\;, \label{sbarbar}
\end{equation}
and $K_{\cal R}(s)$ is the number of sites $i$ for which $i-j$
is odd and $s_i=0$. Note that the definition of $V_j^{\cal R}$
is nothing else but that of $U$, if the sites of the
chain are relabelled as $i\to i-j$.
To illustrate the above definitions, compare the
following examples
\begin{eqnarray}
|+0+-0\rangle\otimes U_5^{\cal R}|-00-++0\rangle&=&
      -|+0+-0\rangle\otimes |+00--+0\rangle\;, \nonumber\\
|+0+-0\rangle\otimes V_5^{\cal R}|-00-++0\rangle&=&
       |+0+-0\rangle\otimes |-00++-0\rangle\;, \nonumber\\
|+0+-0-\rangle\otimes U_6^{\cal R}|00-++0\rangle&=&
      -|+0+-0-\rangle\otimes |00--+0\rangle\;, \nonumber\\
|+0+-0-\rangle\otimes V_6^{\cal R}|00-++0\rangle&=&
      -|+0+-0-\rangle\otimes |00--+0\rangle\;. \nonumber
\end{eqnarray}
It is easy to see that in general $U_j^{\cal R}$ can be
expressed by $V_j^{\cal R}$ as
\begin{equation}
U_j^{\cal R}=p(s) P(s) V_j^{\cal R}\;, \label{UpPV}
\end{equation}
where $p(s)=\pm 1$ is a sign factor and $P(s)$ is either the identity
operator or a general spin-flip $s_i\to -s_i$, $j+1\le i\le N$, on
$\cal R$. To be more specific, let us introduce the notation $Q_{\cal
L}(s)$ for the number of sites $i$ $(i\in \cal L)$ for which $s_i=\pm 1$,
and similarly introduce $Q_{\cal R}(s)$ for $i\in\cal R$.
For the operator $P$, we simply get
\begin{equation}
P(s)=\left\{ \begin{array}{ll}
        \mbox{identity}\; & \mbox{if}\;\; Q_{\cal L}(s)=\mbox{even}\;, \\
        \mbox{spin-flip}\; & \mbox{if}\;\; Q_{\cal L}(s)=\mbox{odd}\;.
          \end{array} \right. \label{bigP}
\end{equation}
As for the sign factor $p$, it is trivially $+1$ if $j={\rm even}$,
while for $j={\rm odd}$ it is
\begin{eqnarray}
p(s) &=& (-1)^{M_{\cal R}(s)-K_{\cal R}(s)}=(-1)^{M_{\cal R}(s)+K_
                                             {\cal R}(s)}\nonumber\\
  &=& (-1)^{N-j-Q_{\cal R}(s)} =-(-1)^{N-Q_{\cal R}(s)},
                          \quad j=\mbox{odd},       \label{smallp}
\end{eqnarray}
since for odd $j$, $K_{\cal R}(s)$ counts the 0's on even sites and
thus $M_{\cal R}(s)+K_{\cal R}(s)$ is the total number of 0's on
$\cal R$.

Let us consider now the states $\Phi_j^{\beta_j,\alpha_{j+1}}
(\alpha_1,\beta_N)$, $\beta_j,\alpha_{j+1}=\uparrow,\downarrow$,
where the valence bond between sites $j$ and $j+1$ is simply
removed. From Eqs.\ (\ref{phifolle}) and (\ref{Uaction})
\begin{equation}
U\Phi_j^{\beta_j,\alpha_{j+1}}(\alpha_1,\beta_N)=
U_j^{\cal L}\Omega_{\cal L}(\alpha_1,\beta_j)\otimes
U_j^{\cal R}\Omega_{\cal R}(\alpha_{j+1},\beta_N)\;.
\end{equation}
Then, using Eq.\ (\ref{UpPV}) we obtain
\begin{equation}
U\Phi_j^{\beta_j,\alpha_{j+1}}(\alpha_1,\beta_N)=
U_j^{\cal L}\Omega_{\cal L}(\alpha_1,\beta_j)\otimes
pPV_j^{\cal R}\Omega_{\cal R}(\alpha_{j+1},\beta_N)\;.
\end{equation}
The expressions for $U_j^{\cal L}\Omega_{\cal L}(\alpha_1,\beta_j)$
and $V_j^{\cal R}\Omega_{\cal R}(\alpha_{j+1},\beta_N)$ can be
read off directly from Eqs.\ (\ref{Oleft}--\ref{Oright}),
using the fact that
the number of sites in the left part is $L=j$, while it is $L=N-j$ in
the right part. Care has to be taken in the proper account of $p$
and $P$. Remember that these depend on the actual configurations
$|s\rangle$. However, fixing $\alpha_1$ and $\beta_j$, the parity
of $Q_{\cal L}(s)$ is uniquely determined for all the
possible configurations in $\Omega_{\cal L}(\alpha_1,\beta_j)$.
Without any loss of generality we will fix the leftmost
spin-1/2 variable to $\alpha_1=\uparrow$ and suppose that
$N$=even.  Then $Q_{\cal L}(s)$=odd and thus $P$ is a spin-flip
[cf.\ Eq.\ (\ref{bigP})] if and only if $\beta_j=\uparrow$.
Similarly, the parity of $Q_{\cal R}(s)$ is uniquely
determined by $\alpha_{j+1}$ and $\beta_N$. From Eq.\ (\ref{smallp})
we easily get
\begin{equation}
 p=\left\{ \begin{array}{ll}
        1\; & \mbox{if}\;\; j=\mbox{even}\;, \\
        -(-1)^{\delta_{\alpha_{j+1},\beta_N}}\; & \mbox{if}\;\;
        j=\mbox{odd}\;. \end{array} \right.
\end{equation}

Using the above results and the fact that a spin flip transforms
$\Psi_1\to\Psi_3$, $\Psi_2\to\Psi_4$ and vice versa, one
straightforwardly obtains the following relations
\begin{eqnarray}
U\Phi_j^{\uparrow,\uparrow}(\uparrow,\uparrow)&=&
    (-)^j (-3)^{N/2}(1/4)(\Psi^{\cal L}_1-\Psi^{\cal L}_2)
        \otimes(\Psi^{\cal R}_3-\Psi^{\cal R}_4)\;,\\
U\Phi_j^{\uparrow,\downarrow}(\uparrow,\uparrow)&=&
        - (-3)^{N/2}(1/4)(\Psi^{\cal L}_1-\Psi^{\cal L}_2)
        \otimes(\Psi^{\cal R}_1+\Psi^{\cal R}_2)\;,\\
U\Phi_j^{\downarrow,\uparrow}(\uparrow,\uparrow)&=&
          (-3)^{N/2}(1/4)(\Psi^{\cal L}_1+\Psi^{\cal L}_2)
        \otimes(\Psi^{\cal R}_1-\Psi^{\cal R}_2)\;,\\
U\Phi_j^{\downarrow,\downarrow}(\uparrow,\uparrow)&=&
   -(-)^j (-3)^{N/2}(1/4)(\Psi^{\cal L}_1+\Psi^{\cal L}_2)
        \otimes(\Psi^{\cal R}_3+\Psi^{\cal R}_4)\;,\\
U\Phi_j^{\uparrow,\uparrow}(\uparrow,\downarrow)&=&
    (-)^j (-3)^{N/2}(1/4)(\Psi^{\cal L}_1-\Psi^{\cal L}_2)
        \otimes(\Psi^{\cal R}_3+\Psi^{\cal R}_4)\;,\\
U\Phi_j^{\uparrow,\downarrow}(\uparrow,\downarrow)&=&
        - (-3)^{N/2}(1/4)(\Psi^{\cal L}_1-\Psi^{\cal L}_2)
        \otimes(\Psi^{\cal R}_1-\Psi^{\cal R}_2)\;,\\
U\Phi_j^{\downarrow,\uparrow}(\uparrow,\downarrow)&=&
          (-3)^{N/2}(1/4)(\Psi^{\cal L}_1+\Psi^{\cal L}_2)
        \otimes(\Psi^{\cal R}_1+\Psi^{\cal R}_2)\;,\\
U\Phi_j^{\downarrow,\downarrow}(\uparrow,\downarrow)&=&
   -(-)^j (-3)^{N/2}(1/4)(\Psi^{\cal L}_1+\Psi^{\cal L}_2)
        \otimes(\Psi^{\cal R}_3-\Psi^{\cal R}_4)\;,
\end{eqnarray}
where the superscript $\cal L$ ($\cal R$) indicates that the
wavefunction $\Psi^{\cal L}$ ($\Psi^{\cal R}$) refers to the
left (right) part of the chain.

Let us insert now the triplet bond in place of the missing valence
bond, i.e.\ symmetrize with respect to the superscripts of
$\Phi_j^{\beta_j,\alpha_{j+1}}$. Recalling Eqs.\
(\ref{pbond}--\ref{nbond}), we find
\begin{eqnarray}
U\Phi_j^+(\uparrow,\uparrow)&=&
  (-)^j (-3)^{N/2}(1/4)(\Psi^{\cal L}_1
        \otimes\Psi^{\cal R}_3-\Psi^{\cal L}_2\otimes\Psi^{\cal R}_3-
        \Psi^{\cal L}_1\otimes\Psi^{\cal R}_4+
        \Psi^{\cal L}_2\otimes\Psi^{\cal R}_4)\;,\\
U\Phi_j^0(\uparrow,\uparrow)&=&
    (-3)^{N/2}(1/2\sqrt{2})(\Psi^{\cal L}_2\otimes\Psi^{\cal R}_1-
        \Psi^{\cal L}_1\otimes\Psi^{\cal R}_2)
                                \;,\label{eq64}\\
U\Phi_j^-(\uparrow,\uparrow)&=&
 -(-)^j (-3)^{N/2}(1/4)(\Psi^{\cal L}_1\otimes\Psi^{\cal R}_3+
        \Psi^{\cal L}_2\otimes\Psi^{\cal R}_3+
        \Psi^{\cal L}_1\otimes\Psi^{\cal R}_4+
        \Psi^{\cal L}_2\otimes\Psi^{\cal R}_4)\;,\\
U\Phi_j^+(\uparrow,\downarrow)&=&
  (-)^j (-3)^{N/2}(1/4)(\Psi^{\cal L}_1\otimes\Psi^{\cal R}_3-
        \Psi^{\cal L}_2\otimes\Psi^{\cal R}_3+
        \Psi^{\cal L}_1\otimes\Psi^{\cal R}_4-
        \Psi^{\cal L}_2\otimes\Psi^{\cal R}_4)\;,\\
U\Phi_j^0(\uparrow,\downarrow)&=&
    (-3)^{N/2}(1/2\sqrt{2})(\Psi^{\cal L}_2\otimes\Psi^{\cal R}_1+
        \Psi^{\cal L}_1\otimes\Psi^{\cal R}_2)\;,\\
U\Phi_j^-(\uparrow,\downarrow)&=&
 -(-)^j (-3)^{N/2}(1/4)(\Psi^{\cal L}_1\otimes\Psi^{\cal R}_3+
        \Psi^{\cal L}_2\otimes\Psi^{\cal R}_3
        -\Psi^{\cal L}_1\otimes\Psi^{\cal R}_4-
        \Psi^{\cal L}_2\otimes\Psi^{\cal R}_4)\;.
\end{eqnarray}
What we obtained is nothing else but a linear combination of the
simplest domain walls $\Psi^{\cal L}_{\nu}\otimes\Psi^{\cal R}_{\nu'}$,
$\nu\ne\nu'$, between sites $j$ and $j+1$. By virtue of the $SU(2)$
symmetry of $H_{\rm VBS}$ and the open boundary condition,
it is possible to consider some linear combinations of
the above states in order to get the simplest forms on the right hand
sides, e.g.,
\begin{eqnarray}
U(\Phi_j^+(\uparrow,\downarrow)+\Phi_j^+(\uparrow,\uparrow)-
  \Phi_j^-(\uparrow,\downarrow)-\Phi_j^-(\uparrow,\uparrow))&=&
    (-)^j (-3)^{N/2} \Psi^{\cal L}_1\otimes\Psi^{\cal R}_3\;,\\
U(\Phi_j^+(\uparrow,\downarrow)-\Phi_j^+(\uparrow,\uparrow)+
  \Phi_j^-(\uparrow,\downarrow)-\Phi_j^-(\uparrow,\uparrow))&=&
    (-)^j (-3)^{N/2} \Psi^{\cal L}_1\otimes\Psi^{\cal R}_4\;,\\
U(\Phi_j^0(\uparrow,\downarrow)-\Phi_j^0(\uparrow,\uparrow))&=&
          (1/\sqrt{2})(-3)^{N/2} \Psi^{\cal L}_1\otimes\Psi^{\cal R}_2\;.
\end{eqnarray}
Three other similar linear combinations can be composed with
$\Psi^{\cal L}_2\otimes\Psi^{\cal R}_{\nu}$, $\nu=1,3,4$, on the right
side. These final forms clearly demonstrates the solitonic
nature of our trial wave functions. Note that in accordance with
the three degrees of freedom of such a spin-1 soliton, there are
three kinds of domain walls. For $\Psi^{\cal L}_1\otimes
\Psi^{\cal R}_3$ ($\Psi^{\cal L}_2\otimes\Psi^{\cal R}_4$) only
${}_{\tilde{H}}\langle S_i^z\rangle _{\tilde{H}}$
changes sign as we move from the left region to the right,
${}_{\tilde{H}}\langle S^x_i\rangle _{\tilde{H}}$ does not change.
The situation is just the opposite for
$\Psi^{\cal L}_1\otimes\Psi^{\cal R}_2$
($\Psi^{\cal L}_2\otimes\Psi^{\cal R}_1$). Here only
${}_{\tilde{H}}\langle S^x_i\rangle _{\tilde{H}}$ flips. Then for
$\Psi^{\cal L}_1\otimes \Psi^{\cal R}_4$ ($\Psi^{\cal L}_2\otimes
\Psi^{\cal R}_3$) the expectation values of the magnetization in
both directions change sign.

Now it is easy to see how $\Phi_j^0$ [or $S^z_j\Omega$,
recalling Eq.\ (\ref{equiv1})] destroys the hidden order
$O^x_{\rm string}$ (and by symmetry $O^y_{\rm string}$) which was
anticipated in Sec.\ III. For definiteness, we fix the boundary
spins $\alpha_1=\uparrow$, $\beta_N=\uparrow$ (other choices can be
worked out similarly) and consider the expectation value
\begin{equation}
{\langle \Phi_j^0(\uparrow,\uparrow)|\sigma_{n,m}^x|
\Phi_j^0(\uparrow,\uparrow) \rangle  \over
|| \Phi_j^0(\uparrow,\uparrow)||^2 }.
\end{equation}
In the case when $j<n$ or $j>m$, i.e.\ $n$ and $m$ are in the same
domain, the domain wall has no effect and the expectation value is
$4/9$. On the other hand when $n<j<m$, it is straightforward to
obtain, using the Kennedy-Tasaki transformation, the following result
in the thermodynamic limit
\begin{eqnarray}
{\langle \Phi_j^0(\uparrow,\uparrow)|\sigma_{n,m}^x|
\Phi_j^0(\uparrow,\uparrow) \rangle  \over
|| \Phi_j^0(\uparrow,\uparrow)||^2 }&=&
{\langle U\Phi_j^0(\uparrow,\uparrow)|S_n^x S_m^x|
U\Phi_j^0(\uparrow,\uparrow) \rangle  \over
|| U\Phi_j^0(\uparrow,\uparrow)||^2 } \nonumber\\
&=&
{\langle \Psi^{\cal L}_2\otimes\Psi^{\cal R}_1
        -\Psi^{\cal L}_1\otimes\Psi^{\cal R}_2|S_n^x S_m^x|
\Psi^{\cal L}_2\otimes\Psi^{\cal R}_1
        -\Psi^{\cal L}_1\otimes\Psi^{\cal R}_2 \rangle  \over
|| \Psi^{\cal L}_2\otimes\Psi^{\cal R}_1
        -\Psi^{\cal L}_1\otimes\Psi^{\cal R}_2||^2 }=
                                                -{4\over 9},
\end{eqnarray}
where we used Eq.\ (\ref{eq64}) and the asymptotic orthogonality of
the different ground states $|\langle\Psi_{\nu '}|
\Psi_{\nu}\rangle|\to 0$ ($\nu'\ne\nu$) if $N\to\infty$.
In fact, the presence of the
domain wall flips the expectation value of $\sigma_{n,m}^x$.

Finally we show that in this formalism the dispersion relation
of Eq.\ (\ref{disp}) can be obtained in a
very elegant way. We can start from e.g.\ the trial wave function
\begin{equation}
|k\rangle=\sum_{j=1}^{N-1} e^{ikj} |j\rangle\; ,
\end{equation}
with $|j\rangle=\phi_1^1\otimes\phi_1^2\otimes\dots\phi_1^j
\otimes\phi_2^{j+1}\otimes\phi_2^{j+2}\otimes\dots\phi_2^{N}$.
In this form $k$ is a variational parameter rather then a momentum,
because of the open boundary condition. In the thermodynamic limit,
however, the boundary condition should not matter
(although it might bring a constant momentum
shift $q$ in the final result, since we have the freedom to redefine
$|j\rangle$ with e.g.\ an arbitrary phase factor $|j\rangle\to
e^{iqj}|j\rangle$),
and the variational energy as a function of
$k$ is in fact the dispersion of the excitations in this simplest
domain wall approach.

The computation would proceed similarly to that in the previous
section. As we have seen there, the important quantities are
$\langle j|j'\rangle$ and $\langle j|\tilde{H}_{\rm
VBS}|j'\rangle$. However, in this case they are trivial because of
the tensor product form. A straightforward calculation gives
\begin{equation}
\langle j|j'\rangle=(-{1\over 3})^{|j-j'|}\;,
\end{equation}
where we used that $\langle \phi_1|\phi_1\rangle=
\langle \phi_2|\phi_2\rangle=1$ and $\langle \phi_1
|\phi_2\rangle=-1/3$, and in a similarly simple way
\begin{equation}
\langle j|\tilde{H}_{\rm VBS}|j'\rangle=\delta_{j,j'}
\langle \phi_1\otimes\phi_2|\tilde{h}_j|
\phi_1\otimes\phi_2\rangle=\delta_{j,j'}{20\over 27}\;,
\end{equation}
which is easily obtained from the explicit form of the two-site
Hamiltonian $\tilde{h}_j$. Apart from a factor of $3^N/2$ these results
are identical to those in Eqs.\ (\ref{norm}) and (\ref{PhiHPhi}).
Therefore, they also lead to the same dispersion $\epsilon(k)$.

\section{Conclusion}

In summary, we studied the elementary excitations in the
valence-bond point of the $S=1$ bilinear-biquadratic model.
Numerical calculations on finite-size systems  were used to predict
the spectrum in the thermodynamic limit. The lowest-lying excited
states above the $k=0$ singlet ground state form a discrete triplet
branch with a minimum at $k=\pi$. Near this minimum this branch is
separated from the higher-lying scattering continuum.  The energy
needed to excite the lowest $k=0$ excitation was found to be twice
the gap value at $k=\pi$. Similarly, the energy of the next lowest
excitation at $k=\pi$ is three times the gap value. These
excitations belong to the continuum and can be interpreted  as
being composed of two or three $S=1$ elementary excitations.

Comparison with the numerical results show that the separate branch
of excitations can be reasonably described with a trial wave
function, where one singlet bond is replaced by a moving triplet
bond. In the representation where the configurations are given in
terms of the $S^z$ eigenstates of the spins, a triplet bond in the
sea of singlet bonds has a solitonic character. In the dilute
system of $+$ and $-$ spin states there is a single domain wall.
While this feature is hidden  in the usual valence-bond
description, it becomes apparent when the nonlocal Kennedy-Tasaki
transformation is used. We have shown that the approximate wave
functions of the excited states transform into explicit domain
walls in the transformed system.

This research was supported in part by the Hungarian Research Fund
(OTKA) Grant Nos.\ T4473 and 2979. GF was also supported by the
Hungarian Scientific Foundation.

\begin{figure}
\caption{Low-lying eigenvalues of the VBS model plotted vs momentum
$k$, for a chain with $N=16$ sites. Labels denote the total spin
$S_T$ of the states. Dashed line shows the energy of the trial
state with the moving hidden soliton.}
\label{spec14}
\end{figure}

\begin{figure}
\caption{Energy gaps $\Delta_A$, $\Delta_B/2$, and $\Delta_C/3$
plotted vs $1/N$. Dashed lines indicate the suggested large $N$
behavior. $\Box$ shows the energy of the trial wave function
at $k=\pi$.}
\label{fss}
\end{figure}


\begin{references}
\bibitem{Ha}{F.\ D.\ M.\ Haldane,
            Phys.\ Rev.\ Lett.\ {\bf 50}, 1153 (1983);
            Phys.\ Lett.\ {\bf 93A}, 464 (1983).}
\bibitem{Af}{For an overall review see I.\ Affleck,
               J.\ Phys.\ Condensed Matter {\bf 1}, 3047 (1989).}
\bibitem{Af-etal}{I.\ Affleck, T.\ Kennedy, E.\ Lieb, and
                                                    H.\ Tasaki,
               Phys.\ Rev.\ Lett.\ {\bf 59}, 799 (1987);
               Commun.\ Math.\ Phys.\ {\bf 115}, 477 (1988).}
\bibitem{Af-Li}{I.\ Affleck and E.\ H.\ Lieb,
               Lett.\ Math.\ Phys.\ {\bf 12}, 57 (1986).}
\bibitem{Ko}{M.\ Kolb,
           Phys.\ Rev.\ B {\bf 31}, 7494 (1985).}
\bibitem{Fa-So}{G.\ F\'ath and J. S\'olyom,
           Phys.\ Rev.\ B {\bf 47}, 872 (1993).}
\bibitem{Fa-Ta}{L.\ D.\ Faddeev and L.\ A.\ Takhtajan,
            Phys.\ Lett.\ {\bf 85A}, 375 (1981);\\
            L.\ A.\ Takhtajan,
            Phys.\ Lett.\ {\bf 87A}, 479 (1982).}
\bibitem{dN-Ro}{M.\ den Nijs and K.\ Rommelse,
           Phys.\ Rev.\ B {\bf 40}, 4709 (1989).}
\bibitem{hidden}{S.\ M.\ Girvin and D.\ P.\ Arovas,
          Phys.\ Scr.\ {\bf T27}, 156 (1989);
          Y.\ Hatsugai and M. \ Kohmoto,
             Phys.\ Rev.\ B {\bf 44}, 11789 (1991);
          F.\ C.\ Alcaraz and Y.\ Hatsugai,
          Phys.\ Rev.\ B {\bf 46}, 13914 (1992).}
\bibitem{Ke-Ta}{T.\ Kennedy and H.\ Tasaki,
           Phys.\ Rev.\ B {\bf 45}, 304 (1992);
           Commun.\ Math.\ Phys.\ {\bf 147}, 431 (1992).}
\bibitem{El-Mi}{N.\ Elstner and H.-J.\ Mikeska,
            Z.\ Phys.\ B {\bf 89}, 321  (1992). }
\bibitem{Go}{G.\ Gomez-Santos,
            Phys.\ Rev.\ Lett.\ {\bf 63}, 790 (1989).}
\bibitem{Ta}{M.\ Takahashi,
          Phys.\ Rev.\ Lett.\ {\bf 62}, 2313 (1989).}
\bibitem{ArAuHa}{D.\ P.\ Arovas, A.\ Auerbach and F.\ D.\ M.\
          Haldane, Phys.\ Rev.\ Lett.\ {\bf 60}, 531 (1988).}
\bibitem{Kna}{S.\ Knabe, J.\ Stat.\ Phys.\ {\bf 52}, 627 (1988). }

\end{references}
\end{document}